\documentclass{ws-ijgmmp}

\begin{document}

\markboth{Winfried Zimdahl}
{Interactions in the dark sector}

%
\catchline{}{}{}{}{}
%

\title{INTERACTIONS IN THE DARK SECTOR OF THE UNIVERSE}

\author{WINFRIED ZIMDAHL}

\address{Departamento de F\'{\i}sica, Universidade Federal do Esp\'{\i}rito Santo,
Avenida Fernando Ferrari, 514, Campus de Goiabeiras, CEP 29075-910, Vit\'oria, Esp\'{\i}rito Santo, Brazil\\
\email{winfried.zimdahl@pq.cnpq.br}}

\maketitle

\begin{history}
\received{(Day Month Year)}
\revised{(Day Month Year)}
\end{history}

\begin{abstract}
Interactions inside the cosmological dark sector influence the cosmological dynamics.
As a consequence, the future evolution of the Universe may be different from that predicted
by the $\Lambda$CDM model.
We review main features of several recently studied models with
nongravitational couplings between dark matter and dark energy.
\end{abstract}


\section{Introduction}

The observed accelerated expansion of the Universe is usually assumed to have its origin in the
existence of a mysterious component with effectively negative pressure, called dark energy (DE).
Together with another up to now exotic component, dark matter (DM), it dominates the dynamics
of the currently observable Universe, at least if standard general relativity (GR) is assumed
to be valid up to the largest cosmological scales.
The homogeneous and isotropic cosmic background dynamics is governed by Friedmann's equation
\begin{equation}\label{friedmann}
3 \frac{\dot{a}^{2}}{a^{2}}
\equiv
H^{2}
= \frac{8 \pi G}{3} \rho_{m} -
\frac{k}{a ^{2}} + \frac{\Lambda}{3}
\end{equation}
and by the acceleration equation
\begin{equation}\label{dda}
\frac{\ddot{a}}{a} = - 4 \pi G\left(\rho_{m} + 3 p_{m} \right) +
\frac{\Lambda}{3}\ ,
\end{equation}
where $H$ is the Hubble rate and $a$ is the scale factor of the Robertson-Walker metric.
The quantities $\rho_{m}$ and $p_{m}$ denote the energy density of the cosmic matter and the corresponding
pressure, respectively. Cold dark matter (CDM) is characterized by a dynamically negligible matter pressure, i.e., $p_{m} \ll \rho_{m}$. Neglecting $p_m$, the equations (\ref{friedmann}) and (\ref{dda}) constitute
the basis of the preferred cosmological model, the $\Lambda$CDM model, which
does well in fitting
most observational data (see, e.g., the recent results from WMAP 9 \cite{wmap9} and Planck \cite{planck}).
Current observations are consistent with a spatially flat universe with fractions of about 70\% DE, provided by the cosmological constant $\Lambda$, and about
30\% matter (including CDM and baryons). But not only because of the notorious cosmological constant and coincidence problems  (see, e.g. \cite{problems}), there is an ongoing interest in alternative models within
GR itself and beyond it. It is useful to test potential deviations from the ``standard" description in
order to constrain additional parameter sets which quantify these deviations.
Among these alternative approaches there are phenomenological fluid models of the dark sector.
These are straightforward generalizations of the $\Lambda$CDM model as can be seen as follows.
With the definitions $\Lambda \equiv 8\pi\,G\,\rho_{\Lambda}$ and
$p_\Lambda \equiv - \rho_{\Lambda}$, the  cosmological constant
is formally equivalent to a perfect ``fluid" with negative
pressure. Then
\begin{equation}\label{lnegp}
H^{2} = \frac{8 \pi G}{3}\rho - \frac{k}{a ^{2}}
\,,\qquad
\
\frac{\ddot{a}}{a} = - \frac{4 \pi G}{3}\left(\rho + 3 p\right)\,,
\end{equation}
where $\rho = \rho_{m } + \rho_{\Lambda}$ and $p = p_{\Lambda}$.
This analogy has been the starting point for generalized fluid models in which either
the equation of state $p_\Lambda \equiv - \rho_{\Lambda}$ or $\rho_{\Lambda} = $ constant or both
are modified. In the following section we summarize basic relation for the dynamics of perfect fluids.

\section{General Perfect Fluid Dynamics}
\label{general}

Dynamical fluid models are based on the energy-momentum tensor of a
perfect fluid,
\begin{equation}
\label{Ttot}
T^{ab} = \rho\,u^{a}u^{b} + p h^{ab}\,,\quad h^{ab} \equiv g^{ab}
+ u^{a}u^{b}\,,\quad h^{ab}\,u_{b}  = 0\,,
\end{equation}
with $u_{a}u^{a} = - 1$ and $u_{a}h^{ab} = 0$.
Local energy conservation is equivalent to
\begin{equation}\label{}\
u_{a}T^{ab}_{\ ;b} = 0 \quad \Rightarrow \quad \dot{\rho} + \Theta\,\left(\rho + p\right)  = 0\,,
\end{equation}
where the expansion scalar $\Theta \equiv u^{a}_{;a}$  reduces to $3H$ in the homogeneous and isotropic background.
Projection orthogonal to $u_{a}$ yields the momentum conservation
\begin{equation}\label{}
h_{a}^{m}\,T^{ab}_{\ ;b} = \left(\rho + p\right)\,\dot{u}^{m} + p_{,b}h^{mb}   = 0\,.
\end{equation}
The spatial projection of the covariant derivative
$u_{m;b}$ may be decomposed according to
\begin{equation}\label{}
h_{m}^{a}h_{b}^{c}u_{a;c} = \omega_{mb}  + \sigma_{mb} + \frac{1}{3}\Theta h_{mb}
\end{equation}
with the antisymmetric and symmetric trace-free parts
\begin{equation}\label{}
\omega_{ab} = h_{a}^{c}h_{b}^{d}u_{\left[c;d\right]}\ \quad \mathrm{and}\quad
\sigma_{ab} = h_{a}^{c}h_{b}^{d}u_{\left(c;d\right)}
- \frac{1}{3}\Theta_{}h_{ ab}\,,
\end{equation}
respectively. The time evolution of the expansion scalar is governed by
Raychaudhuri's equation
\begin{equation}\label{}
\dot{\Theta} + \frac{1}{3}\Theta^{2} - 2 \left(\omega^{2} -
\sigma^{2}\right) - \dot{u}^{a}_{;a} + 4\pi G \left(\rho + 3
p\right) = 0\,,
\end{equation}
where $\sigma^{2}$ and $\omega^{2}$ are the scalars
\begin{equation}
\sigma^{2} = \frac{1}{2}\sigma_{ab}\sigma^{ab}\quad \mathrm{and} \quad
\omega^{2} = \frac{1}{2}\omega_{ab}\omega^{ab}\  \label{}
\end{equation}
of shear and vorticity, respectively. For the Friedmann-Lema\^{\i}tre-Robertson-Walker (FLRW) cosmological models $\sigma^{2} = 0$ and $\omega^{2} = 0$ are valid.
In the following section we generalize the one-component description to the case of two coupled fluids.

\section{Interacting Fluids}

\subsection{General relations}

For a two-component system, the total energy-momentum tensor (\ref{Ttot})
is split into a matter part $T_{m}^{ik}$ and a part $T_{x}^{ik}$ which is supposed to describe a dynamical form of DE,
\begin{equation}\label{}
T^{ik} = T_{m}^{ik} + T_{x}^{ik}\,.
\end{equation}
For both parts we assume a perfect-fluid structure, i.e.,
\begin{equation}\label{}
T_{A}^{ik} = \rho_{A} u_A^{i} u^{k}_{A} + p_{A} h_{A}^{ik} \
,\qquad\ h_{A}^{ik} = g^{ik} + u_A^{i} u^{k}_{A}\ , \qquad A = m, x\,.
\end{equation}
For separately conserved components $T_{m}^{ik}$ and $T_{x}^{ik}$, the $\Lambda$CDM model can be seen as a special case with $\rho_{x} = \rho_{\Lambda} =$ constant and
$p_{m} = 0$.
Generally, total energy-momentum conservation $T_{\ ;k}^{ik} = 0$ is compatible with a coupling between both components,
\begin{equation}\label{}
T_{m\ ;k}^{ik} = Q^{i},\qquad T_{x\ ;k}^{ik} = - Q^{i}\,,
\end{equation}
where the quantity $Q^{i}$ appears as a source (or sink) in the individual balance equations.
The separate energy-balance equations are
\begin{equation}
-u_{mi}T^{ik}_{m\ ;k} = \rho_{m,a}u_{m}^{a} +  \Theta_{m}\rho_{m} = -u_{ma}Q^{a}\
\label{eb1}
\end{equation}
and
\begin{equation}
-u_{xi}T^{ik}_{x\ ;k} = \rho_{x,a}u_{x}^{a} +  \Theta_{x} \left(\rho_{x} + p_{x}\right) = u_{xa}Q^{a}\ .
\label{eb2}
\end{equation}
In general, each component has its own four-velocity $u_{A}^{i}$ with $g_{ik}u_{A}^{i}u_{A}^{k} = -1$. The rates $\Theta_{A}$ are defined as $\Theta_{A} = u^{a}_{A;a}$. For the background dynamics we assume all four-velocities to coincide, i.e. $u_{m}^{a} = u_{x}^{a} = u^{a}$. For the momentum balances it follows that
\begin{equation}
h_{mi}^{a}T^{ik}_{m\ ;k} = \rho_{m} \dot{u}_{m}^{a} = h_{mi}^{a} Q^{i}\
\label{mb1}
\end{equation}
and
\begin{equation}
h_{xi}^{a}T^{ik}_{x\ ;k} = \left(\rho_{x} + p_{x}\right)\dot{u}_{x}^{a} + p_{x,i}h_{X}^{ai} = - h_{x i}^{a} Q^{i}\,,
\label{mb2}
\end{equation}
where $\dot{u}_{A}^{a} \equiv u_{A ;b}^{a}u_{A}^{b}$.
Equation (\ref{mb1}) implies that in the presence of a coupling term the CDM fluid motion is nongeodesic in general.

The interaction term $Q^{i}$ can be split into parts proportional and perpendicular to the total four-velocity according to
\begin{equation}
Q^{i} = u^{i}Q + \bar{Q}^{i}\ ,
\label{Qdec}
\end{equation}
where $Q = - u_{i}Q^{i}$ and $\bar{Q}^{i} = h^{i}_{a}Q^{a}$, with $u_{i}\bar{Q}^{i} = 0$.
Alternatively, a similar split with respect to the matter four-velocity may be useful.

\subsection{Background dynamics}

In the homogeneous and isotropic background the set of equations (\ref{eb1}) and (\ref{eb2}) reduces to
\begin{equation}\label{balbg}
\dot{\rho}_{m} + 3H \rho_{m} = Q \ ,\quad \dot{\rho}_{x} +
3H (1+w)\rho_{x} = - Q\,,
\end{equation}
where $w \equiv p_{x}/\rho_{x}$ is the equation-of-state (EoS) parameter of the DE component.
Equations (\ref{mb1}) and (\ref{mb2}) are satisfied identically.

Straightforwardly one realizes that the source (loss) term $Q$ does not directly enter the Hubble rate and the deceleration parameter $q = - 1 - \dot{H}/H^{2}$ which, aside from $H$, is determined by its derivative $\dot{H}$.
The lowest order at which $Q$ appears explicitly is in the second derivative of the Hubble rate \cite{ZP},
\begin{equation}\label{ddH}
\frac{\ddot{H}}{H^{3}} = \frac{9}{2} +
\frac{9}{2}w\frac{\rho_x}{\rho}\left[2 + w +
\frac{1}{3H}\left(\frac{Q}{\rho_{x}} -
\frac{\dot{w}}{w}\right)\right]\ .
\end{equation}
The influence of the interaction on the dynamics may be quantified by the
statefinder parameter (``jerk'') \cite{chiba,staro,chenalcaniz}
\begin{equation}\label{j}
j\equiv \frac{1}{aH^3}\frac{\mbox{d}^3 a}{\mbox{d}t^3} = 1 +
3\frac{\dot H}{H^2} + \frac{\ddot H}{H^3}\ .
\end{equation}
The parameter $j$ enters the luminosity distance
\begin{equation}\label{dL}
d_L =
\left(1+z\right)\int \frac{\mbox{d}z}{H\left(z\right)}
\end{equation}
in third order in the redshift \cite{statef}:
\begin{equation}\label{dL3}
d_L \approx  \frac{z}{H_{0}} \left[1 + \frac{1}{2}\left(1 -q_0
\right)z + \frac{1}{6}\left( 3\left(q_0 + 1\right)^2 - 5\left(q_0
+ 1\right) + 1 - j_0 \right)z^2\right].
\end{equation}
The subscript 0 denotes the present value of the corresponding quantity.
While for the $\Lambda$CDM model $j = j_{0} = 1$ is valid, coupled models have $j_{0} \neq 1$ in general.
Potentially, this allows us to discriminate between models that share the same values of $H_{0}$ and $q_{0}$.

\subsection{Perturbations}

\begin{figure}[ph]
\centerline{\psfig{file=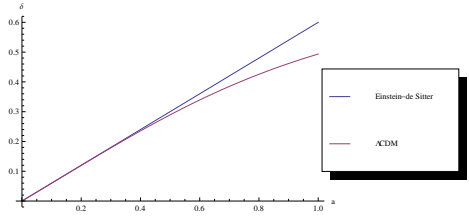,width=2.5in}}
\vspace*{8pt}
\caption{Growth rates for the $\Lambda$CDM model and the Einstein-de Sitter universe.\label{growth}}
\end{figure}

To describe inhomogeneities, we perform a split of all variables into a homogeneous part
and first-order perturbations about the homogeneous and isotropic background.
First-order fluid perturbation variables will be denoted by a hat symbol on top of the respective quantity.
Of particular importance are the perturbations of the matter energy density
for which we define the fractional
perturbation $\delta_{m}$:
\begin{equation}\label{}
\rho_{m} \Rightarrow\ \rho_{m}(t) + \hat{\rho}_{m}(\mathbf{x},t)\ ,\quad \delta_{m} \equiv \frac{\hat{\rho}_{m}}{\rho_{m}}\,.
\end{equation}
With the definitions
\begin{equation}\label{}
\Omega_{m0} = \frac{8\pi G \rho_{m0}}{3 H^{2}_{0}}\, \quad \mathrm{and} \quad
\Omega_{\Lambda} = \frac{\Lambda}{3 H^{2}_{0}}\,,
\end{equation}
the equation for $\delta_{m}$ for the $\Lambda$CDM model is
\begin{equation}\label{deltaalcdm}
\delta_{m}''+\frac{3}{a}\left[ 1-\frac{\Omega_{m0}a^{-3}}{2(\Omega_{m0}a^{-3}+\Omega_{\Lambda})}\right]
 \delta_{m}'-\frac{3}{2a^{2}}\frac{\Omega_{m0}a^{-3}}{(\Omega_{m0}a^{-3}+\Omega_{\Lambda})}\delta_{m} =0\,, \quad  (\Lambda \mathrm{CDM})\,,
\end{equation}
where the prime means a derivative with respect to $a$.
Equation (\ref{deltaalcdm}) generalizes the corresponding equation
\begin{equation}\label{deltaEdS}
\delta_{m}''+\frac{3}{2 a}\delta_{m}'-\frac{3}{2a^{2}}\delta_{m}=0\,, \qquad \qquad (\mathrm{Einstein-de\ Sitter})\,,
\end{equation}
for the Einstein-de Sitter universe. The latter is recovered from (\ref{deltaalcdm}) in the
limit $\Omega_{\Lambda} = 0$.  Equation~(\ref{deltaEdS}) has a solution $\delta_{m} \propto a$, which describes the growth of density perturbations in the matter-dominated era.
The behavior of $\delta_{m}$ for a typical perturbation for both (\ref{deltaalcdm}) and (\ref{deltaEdS}) is visualized in Fig.~\ref{growth}. Obviously, the existence of DE attenuates the growth of matter perturbations. Different DE models, in particular coupled models, will predict a modified growth of $\delta_{m}$. Different growth rates may serve to remove degeneracies for models which otherwise share the same background dynamics.

The two-component linear perturbation dynamics is intrinsically nonadiabatic. The relevant combination is
\begin{equation}\label{}
\hat{p}- \frac{\dot{p}}{\dot{\rho}}\hat{\rho} = \hat{p}_{x} -\frac{\dot{p}_{x}}{\dot{\rho}}\,\hat{\rho}
= \hat{p}_{x} + \frac{\dot{\rho}_{x}}{\dot{\rho}}\left(\hat{\rho}_{m} + \hat{\rho}_{x}\right)\,.
\end{equation}
Adiabatic perturbations $\hat{p}_{ad}$ are characterized by a vanishing of this combination,
\begin{equation}\label{}
\hat{p}_{ad} - \frac{\dot{p}}{\dot{\rho}}\hat{\rho} = 0\,.
\end{equation}
The nonadiabatic part of the pressure perturbations turns out to be
\begin{equation}\label{}
\hat{p}_{nad} \equiv\hat{p}- \frac{\dot{p}}{\dot{\rho}}\hat{\rho} = \frac{\dot{\rho}_{x}\dot{\rho}_{m}}{\dot{\rho}}
\left(\frac{\hat{\rho}_{m}}{\dot{\rho}_{m}} - \frac{\hat{\rho}_{x}}{\dot{\rho}_{x}}\right)\,.
\end{equation}
Here we have assumed that the DE component is adiabatic on its own.
Nonadiabatic perturbations are present in all the models to be discussed in the following, although
at high redshift the nonadiabaticity may be quantitatively negligible.

Restricting ourselves to scalar perturbations, the line element can be written
\begin{equation}\label{line}
\mbox{d}s^{2} = - \left(1 + 2 \phi\right)\mbox{d}t^2 + 2 a^2
F_{,\alpha }\mbox{d}t\mbox{d}x^{\alpha} +
a^2\left[\left(1-2\psi\right)\delta _{\alpha \beta} + 2E_{,\alpha
\beta} \right] \mbox{d}x^\alpha\mbox{d}x^\beta\
\end{equation}
with the scalar perturbation variables $\phi$, $F$, $\psi$ and $E$.
For the spatial components $\hat{u}^\mu$ of the four-velocity we introduce a velocity potential $v$ by \cite{vdf}
\begin{equation}\label{v}
a^2\hat{u}^\mu + a^2F_{,\mu} = \hat{u}_\mu \equiv v_{,\mu}\ .
\end{equation}
None of the first-order perturbation quantities is invariant under infinitesimal coordinate transformations.
The perturbation dynamics for the models to be studied below will be formulated in terms of the gauge-invariant
combinations
\begin{equation}\label{invc}
\delta_{m}^{c} \equiv \delta_{m} + \frac{\dot{\rho}_{m}}{\rho_{m}} v\,,\quad  \delta_{x}^{c} \equiv \delta_{x} + \frac{\dot{\rho}_{x}}{\rho_{x}} v\,, \quad \hat{p}_{x}^{c} \equiv \hat{p}_{x} + \dot{p}_{x} v \,,\quad
\hat{\Theta}^{c} \equiv \hat{\Theta} + \dot{\Theta} v \,.
\end{equation}
The superscript c indicates that the corresponding quantity acquires its physical meaning in the comoving frame
$v=0$.

For the comparison with observational data one calculates, e.g., the matter power spectrum
\begin{equation}\label{Pk}
P_k=\left|\delta_{m,k}^{c}\right|^{2}\,,
\end{equation}
where $\delta_{m,k}^{c}$ is the Fourier component of the density contrast $\delta_{m}^{c}$. To test a given model with a set $\left\{\mathbf{p}\right\}$ of free parameters,
one minimizes the quantity
\begin{equation}\label{chi2}
\chi^{2}\left({\bf p}\right)=\frac{1}{N_f}\sum_{i}\frac{\left[P^{th}_{i}({\bf p}) - P^{obs}_{i}({\bf p})\right]^{2}}{\sigma_{i}^{2}}\,.
\end{equation}
Here,  $N_f$ is the number of degrees of freedom, $P^{th}_{i}$ and
$P^{obs}_{i}$ are the theoretical and the observed values, respectively, for the power spectrum and
$\sigma_{i}$ denotes the error for the data point $i$.
The statistical analysis for other data sets, below it will be relevant also for data from supernovae of type Ia, is done in a similar way.

The relations so far did not specify any interaction and are generally valid. In the following we shall
review some recently studied specific models, all of them on the basis of GR.
Since neither the physical nature of DE nor that of DM are known, there is no real guiding principle for the choice of specific interactions in the dark sector either. Therefore, all considerations are purely phenomenological and explore potential consequences of different hypothetical couplings between DM and DE.

\section{Models of Interacting Dark Energy}

\subsection{Scaling cosmology}
This approach is based on a phenomenological ansatz for the dynamics of the
energy-density ratio $r$ of DM and DE \cite{dalal},
\begin{equation}\label{rxi}
r = \frac{\rho_m}{\rho_x} = r_{0}a^{-\xi}\,,
\end{equation}
where $\xi$ is the scaling parameter and $r_{0}$ is the present value of the ratio $r$.
The $\Lambda$CDM model is recovered for $\xi = 3$ together with $w=-1$.
A stationary ratio has $\xi =  0$.
According to \cite{dalal}, one may quantify the severity of the coincidence problem by the phenomenological parameter $\xi$. In this sense, any $\xi < 3$ alleviates this problem.

\begin{figure}[ph]
\ \\
\ \\
\centerline{\psfig{file=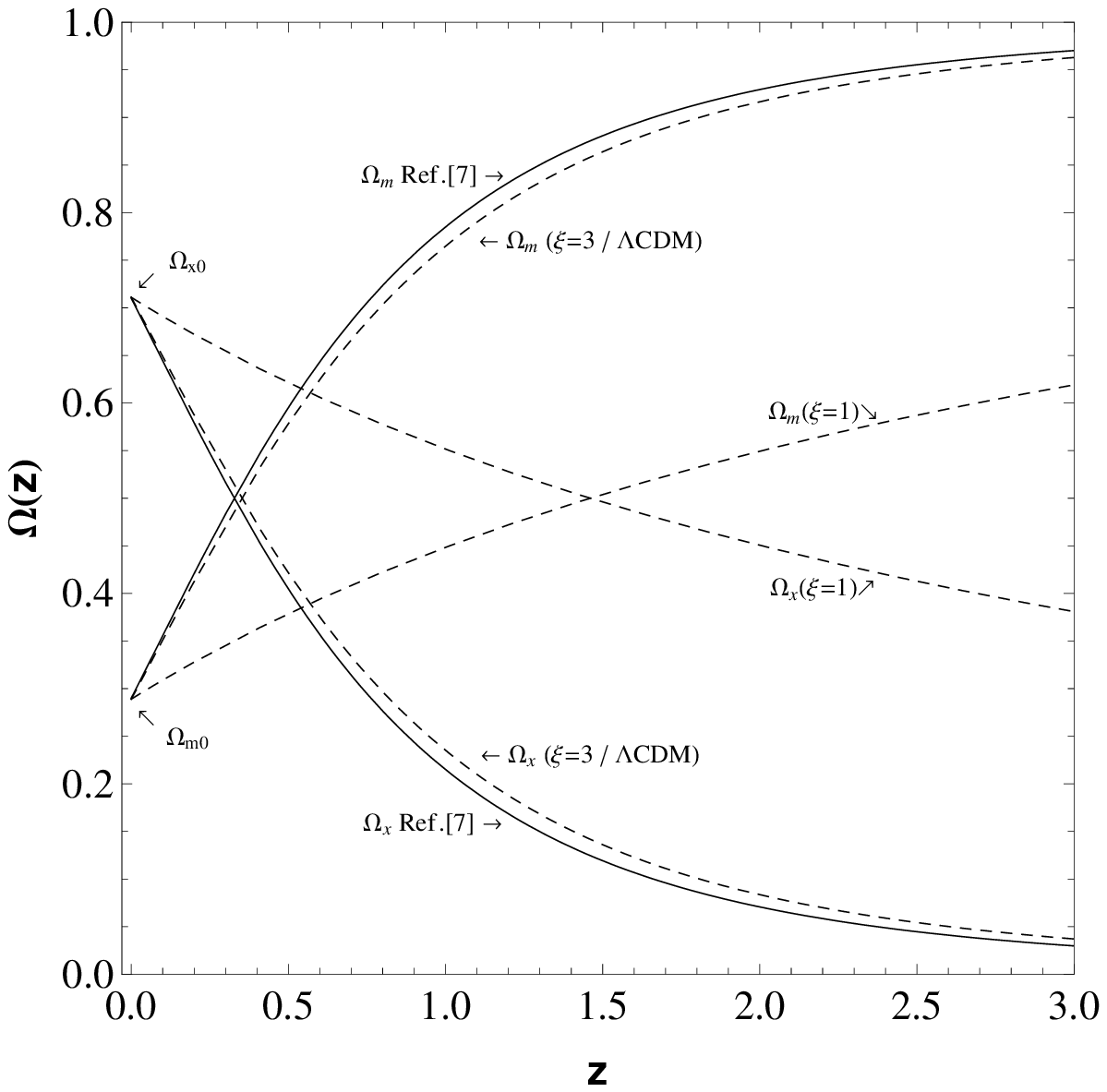,width=1.7in}}
\vspace*{8pt}
\caption{
Redshift dependence of the fractional contributions $\Omega_{m}$ and $\Omega_{x}$
with the same values $\Omega_{m0}=0.289$ and $\omega = -1.01$ (cf.~[7]) for all models.
While the curves for the interacting model of [7] (solid lines) and for our $\xi=3$ model  (which is indistinguishable from the $\Lambda$CDM model) are similar to each other, the difference between $\Omega_{m}$ and $\Omega_{x}$ is much smaller for $\xi=1$
than for any of the other models.
\label{coincidence}}
\end{figure}

Combining the energy balances (\ref{balbg}) with the ansatz (\ref{rxi}) allows us to obtain an expression for $Q$ in terms
of the EoS parameter $w$ and the scaling parameter $\xi$ \cite{ZP},
\begin{equation}\label{Qscale}
Q = - 3 H\, \frac{\frac{\xi }{3}
+ w} {1 + r_{0} \left(1+z \right)^{\xi }}\, \rho _{m}\,.
\end{equation}
Consequently, given a value of $w$, a suitable interaction is required to
produce a certain scaling behavior of the type (\ref{rxi}).
A stationary ratio, in particular, is characterized by a power-law behavior:
\begin{equation}\label{}
r = r_0 = {\rm const}\quad \Rightarrow\quad \rho _{x} \ ,\ \rho
_{m} \propto a ^{-\nu }\,,\quad \nu = 3 \frac{1 + r_0 + w}{1 + r_0}\ , \quad a \propto
t^{2/\nu}\,.
\end{equation}
The condition to have accelerated expansion is $ \nu < 2\, \leftrightarrow 3w < - \left
(1+r_0\right)$.
This dynamics has a scalar field representation with an exponential potential
$V(\phi) \propto\ exp{\left[- \lambda \phi \right]}$.
The condition for accelerated expansion translates into a condition for the parameter $\lambda$ \cite{ZPC},
\begin{equation}\label{}
\ddot{a} > 0 \quad \Leftrightarrow\quad \lambda ^{2} < 24\pi G \frac{w^{2}}{\left(1+r_0 \right)
\left(1+w\right)}\,.
\end{equation}
With the definitions
\begin{equation}\label{}
\rho_{cr}\equiv \frac{3 H^{2}}{8\pi G}\,,\quad \Omega_{m} \equiv \frac{\rho_{m}}{\rho_{cr}}\,,\quad
\Omega_{x} \equiv \frac{\rho_{x}}{\rho_{cr}},
\end{equation}
the potential alleviation of the coincidence problem for $\xi < 3$ is visualized in Fig.~\ref{coincidence} (see also \cite{chenalcaniz}).
For $\xi=1$, e.g., the values of $\Omega_{m}$ and $\Omega_{x}$ are much closer to each other over a certain redshift range than for the
$\Lambda$CDM model with $\xi=3$.

The scaling dynamics may be generalized to
variable equations of state and to include a baryonic component (subscript b) \cite{david}. Using the popular
CPL \cite{cpl} parametrization $w = w_{0} + w_{1}\left(1 - a\right)$, the special case $\xi = 1$
admits an analytic solution for the
Hubble rate,
\begin{equation}\label{}
\left[ \frac{H(a)}{H_0}\right] ^2=\frac{\Omega_{b0}}{a^3}+ \frac{\left( 1-\Omega_{b0}\right)^{\left(1+3y \right)}}{\left(1-\Omega_{b0}+\Omega_{m0}z\right)^{3y}a^{3\left(1+y  \right)}}
\exp\left[3w_1 \left(a-1\right)\right]
\end{equation}
with
\begin{equation}\label{}
z = \frac{1}{a} -1\,, \qquad
y \equiv w_0+w_1\left(\frac{1-\Omega_{b0}}{1-\Omega_{b0}-\Omega_{m0}}\right)\,,
\end{equation}
where $\Omega_{b0} = 8\pi G \rho_{b0}/(3 H^{2}_{0})$.
The results of a statistical analysis (similar to (\ref{chi2})), based on the Union2 data set \cite{union2}, for the cases $\xi = 1$ and $\xi = 3$ together with a comparison
with the $\Lambda$CDM model are summarized in Table~1.
The AIC and BIC criteria in the last two lines take into account the number of degrees of freedom of the model under consideration.
The AIC criterion uses the formula $\mathrm{AIC} = \chi^2_{min} + 2k$ \cite{akaike}, where $k$ is the number of degrees of freedom. The
BIC criterion \cite{schwarz} is based on the expression $\mathrm{BIC} = \chi^2_{min} + 2k\ln N$, where $N$ is the number of observational points. The smaller the resulting numbers in both expressions, the higher the quality of the corresponding model.
Different models are classified with respect to the differences $\Delta$AIC and $\Delta$BIC between its AIC and BIC values, respectively, and the corresponding values for a reference model. This establishes a scale which allows for a ranking of different models according to the magnitude of their differences  $\Delta$AIC and $\Delta$BIC (see, e.g., \cite{liddle}).
Notice that the $\chi^{2}_{min}$ value for $\xi = 3$ is smaller than that of the $\Lambda$CDM model. But the mentioned criteria penalize the introduction of additional parameters and reverse the ranking.
This is a typical feature for many alternative models.
Using the AIC criterion, the $\xi = 1$ and $\xi = 3$ models can be considered as still weakly supported ($\Delta$AIC $< 6$).
On the basis of the BIC criterion, however, these models are disfavored ($\Delta$BIC $> 10$).
This kind of contradiction in using
different evaluation criteria is well known in the literature, see, e.g., \cite{syz}.
In any case, the $\Lambda$CDM model is the clear winner of the competition and this way of alleviating  the coincidence problem does not seem to be supported by the data.

\begin{table}[ht]
\tbl{Summary of the analysis for the Union2 data set (557 supernovae).}
{\begin{tabular}{@{}cccc@{}} \toprule
Model & $\Lambda$CDM & $\xi=1$ & $\xi=3$  \\
\colrule
Best fit\hphantom{00} & \hphantom{0}$\Omega_{m0} = 0.268$ & \hphantom{0}$\Omega_{m0} = 0.270$ & $\Omega_{m0} =0.272 $ \\
\hphantom{00} & \hphantom{0}& \hphantom{0}$w_0 = -1.081$ & $w_0 = -1.018$\\
\hphantom{0} & & $ \hphantom{0} w_1 = 1.269$ &  $w_1 =0.092 $\hphantom{0} \\
\colrule
$q(z=0)$ & $q_0=-0.598$ & $q_0=-0.683$\hphantom{0} & $q_0=-0.702$\\
\colrule
$\chi^2_{min}$ & 541.156 & 541.300 \hphantom{0} & 540.997  \\
$k$ & 1& 3\hphantom{0} & 3 \\
$\Delta$BIC  & 0& 12.789 \hphantom{0} & 12.486  \\
$\Delta$AIC  & 0 & 4.144  \hphantom{0} &  3.841  \\
\botrule
\end{tabular}}
\end{table}

\subsection{Transient acceleration}

The idea that the currently observed accelerated expansion of the Universe might be
a transient phenomenon has been discussed several times in the literature
(\cite{albrecht,barrow,bertolami,alcaniz,sastaro,antonio}).
Here, we describe a model for which such type of behavior is the consequence of an interaction in the dark sector \cite{nelson,cristofherpert}. Generally, an interaction modifies the $a^{-3}$ behavior of the
matter energy density to
\begin{equation}\label{rhommod}
\rho_{m} =
\rho_{m0}a^{-3}\,\tilde{f}(a)\,,\quad \Rightarrow\quad Q = \frac{\dot{\tilde{f}}}{\tilde{f}}\rho_{m}\,,
\end{equation}
where the function $\tilde{f}(a)$ encodes the influence of the interaction.
Then, the corresponding DE balance can be written in terms of an effective EoS parameter $w^{eff}$,
\begin{equation}\label{ebalhol}
\dot{\rho}_{x} = - 3H \left(1 + w^{eff}\right)\rho_{x}, \quad
w^{eff} = w + \frac{\dot{\tilde{f}}}{3 H \tilde{f}}\, r \ .
\end{equation}
It is useful to parametrize the interaction according to
\begin{equation}\label{fg}
\tilde{f}\left(a\right)=1+g\left(a\right)
\end{equation}
and to consider the special case
\begin{equation}\label{gauss}
w = -1 \ , \quad g(a)=\gamma\, a^5 \exp (-a^2/\sigma ^2)\ ,
\end{equation}
where $\gamma$ is an interaction constant.
Although the expression for $g(a)$ in (\ref{gauss}) was taken for mathematical convenience,
it admits an analytic solution of the background dynamics, it can serve to demonstrate some
general features for interaction-induced transient acceleration.

Integration of the DE balance (\ref{ebalhol}) yields
\begin{equation}\label{rhoxtrans}
\rho_{x} = \rho_{x_{0}}^{eff}
- \gamma\,\frac{\rho_{m_0}}{1+g_{0}}\,\exp\left(-a^{2}/\sigma^{2}\right) \left(a^{2} - \frac{3}{2}\sigma^{2}\right)\,,
\end{equation}
where $\rho_{x_{0}}^{eff}$ plays the role of an effective cosmological constant,
\begin{equation}\label{effconst}
\rho_{x_{0}}^{eff}
= \rho_{x_{0}} - \frac{3}{2}\gamma\,\frac{\rho_{m_0}}{1+g_{0}}\,\exp (-1/\sigma^{2})
\left[\sigma^{2} - \frac{2}{3}\right]\ .
\end{equation}
Obviously, a transient acceleration is only possible if $\rho_{x_{0}}^{eff}=0$. Otherwise, the constant $\rho_{x_{0}}$
would always prevail in the long-time limit.
Consequently, for accelerated expansion to be a transient phenomenon, part of the interaction has to cancel the bare cosmological constant $\rho_{x_{0}}$.
Under this condition the acceleration equation becomes
\begin{equation}
\frac{\ddot a}{a}= -\frac{H_{0}^{2}}{2}\left\{
\left[\frac{1 - \frac{3}{2}K\sigma^{2}
\exp
(-1/\sigma^{2})}{a^{3}}\right]
-
3 K \exp
(-a^{2}/\sigma^{2}) \left[\sigma^{2} -a^{2}\right]\right\}\,,
\label{ddatrans}
\end{equation}
where
\begin{equation}\label{defK}
 K = \frac{8 \pi G}{3 H_{0}^{2}}\gamma\frac{\rho_{m_0}}{1+g_{0}}\,
\end{equation}
quantifies the interaction.
It is useful to compare (\ref{ddatrans}) with the corresponding expression for the $\Lambda$CDM model,
\begin{equation}\label{ddalcdm}
\frac{\ddot a}{a}=-\frac{H_{0}^{2}}{2}
\left\{\frac{1 - \Omega_{\Lambda}}{a^{3}}
- 2 \Omega_{\Lambda}\right\}\,,\qquad\qquad (\Lambda \mathrm{CDM})\ .
\end{equation}
Our alternative model is not expected to deviate too strongly from the $\Lambda$CDM model at the present time.
A comparison between (\ref{ddatrans}) and (\ref{ddalcdm}) then suggests that $K$ should be positive and that the interaction term plays a similar role as the cosmological constant $\Lambda$. In other words, the role of the interaction is twofold. As already mentioned, it has to cancel the bare cosmological constant. But at the same time it has to induce an accelerated expansion by itself.

\begin{figure}[ph]
\centerline{\psfig{file=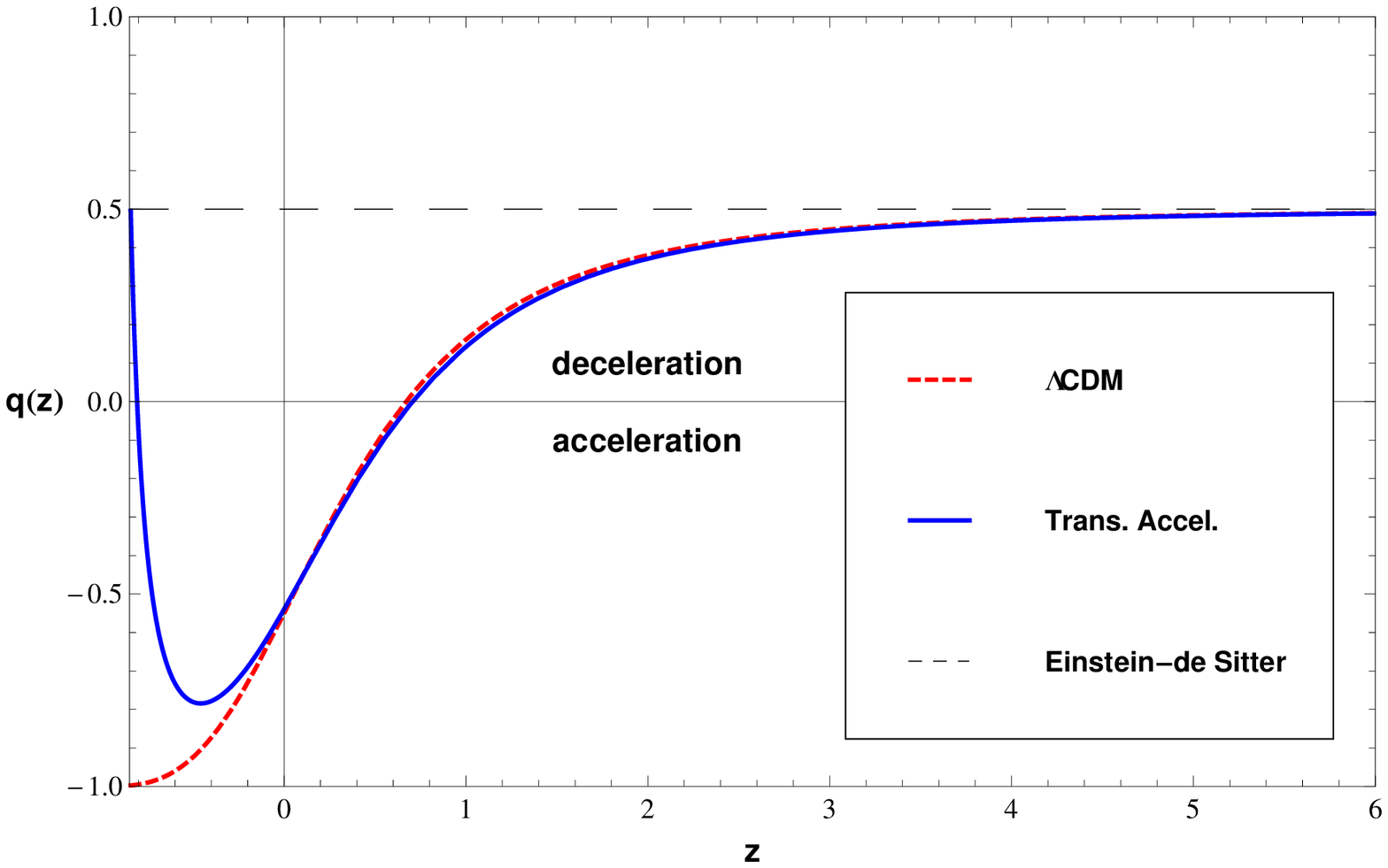,width=2.7in}}
\vspace*{8pt}
\caption{The deceleration parameter of the transient acceleration model as function of the redshift
for the best-fit parameters (solid line). The dashed line shows the corresponding dependence for the $\Lambda$CDM model. The value $q= 0.5$ corresponds to the Einstein-de Sitter universe.\label{figtrans}}
\end{figure}

The expression (\ref{ddatrans}) implies an
early ($a\ll 1$) decelerated expansion for $\Omega_{m_{0}} > K\exp (-1/\sigma^{2})$ which represents an
upper limit on the interaction strength $K$.
The presently observed accelerated expansion
corresponds to the condition
\begin{equation}\label{K>}
\frac{\ddot{a}}{aH^{2}}\mid _{0} \ > 0 \quad \Leftrightarrow \quad
K\,\exp (-1/\sigma^{2})\left[\sigma^{2} - \frac{2}{3}\right] > \frac{2}{9}\ ,
\end{equation}
equivalent to a lower limit on the interaction strength.
This means, there is an admissible
range
\begin{equation}\label{range}
\frac{2}{9}\frac{e^{1/\sigma^{2}}}{\sigma^{2} - \frac{2}{3}} < K <
\frac{2 e^{1/\sigma^{2}}}{3\sigma^{2}}
\end{equation}
for $K$. Using the Constitution data \cite{constitution},
Fig.~\ref{figtrans} shows that this model indeed describes an early transition
from decelerated to accelerated expansion together with a future transition back to decelerated expansion \cite{cristofherpert}.
Table~2 summarizes the best-fit values of the parameters $\sigma$, $K$ and $h$.
\begin{table}[ht]
\tbl{Best-fit values, based on the
Constitution data (397 supernovae),  for the parameters $\sigma$, $K$ and $h$.}
{\begin{tabular}{@{}cccc@{}} \toprule
\hphantom{00}$\chi^{2}_{\mbox{\tiny{min}}}$ \hphantom{00}
& \hphantom{0}$\sigma$  & $K$  & $h$ \\
\colrule
$465.5$ \hphantom{00} & \hphantom{0}$5.23 ^{+0.05}_{-0.05}$ & \hphantom{0}$0.018 ^{+0.0004}_{-0.0004} $ &
$0.65 ^{+0.003}_{-0.003}$\\
\botrule
\end{tabular}}
\end{table}
While the choice of the interaction (\ref{gauss})
may seem to be tailored to produce the expected behavior, it is not trivial that there exists a range
(\ref{range}) which is compatible with current observational data.

\begin{figure}[ph]
\ \\
\ \\
\ \\
\ \\
\centerline{\psfig{file=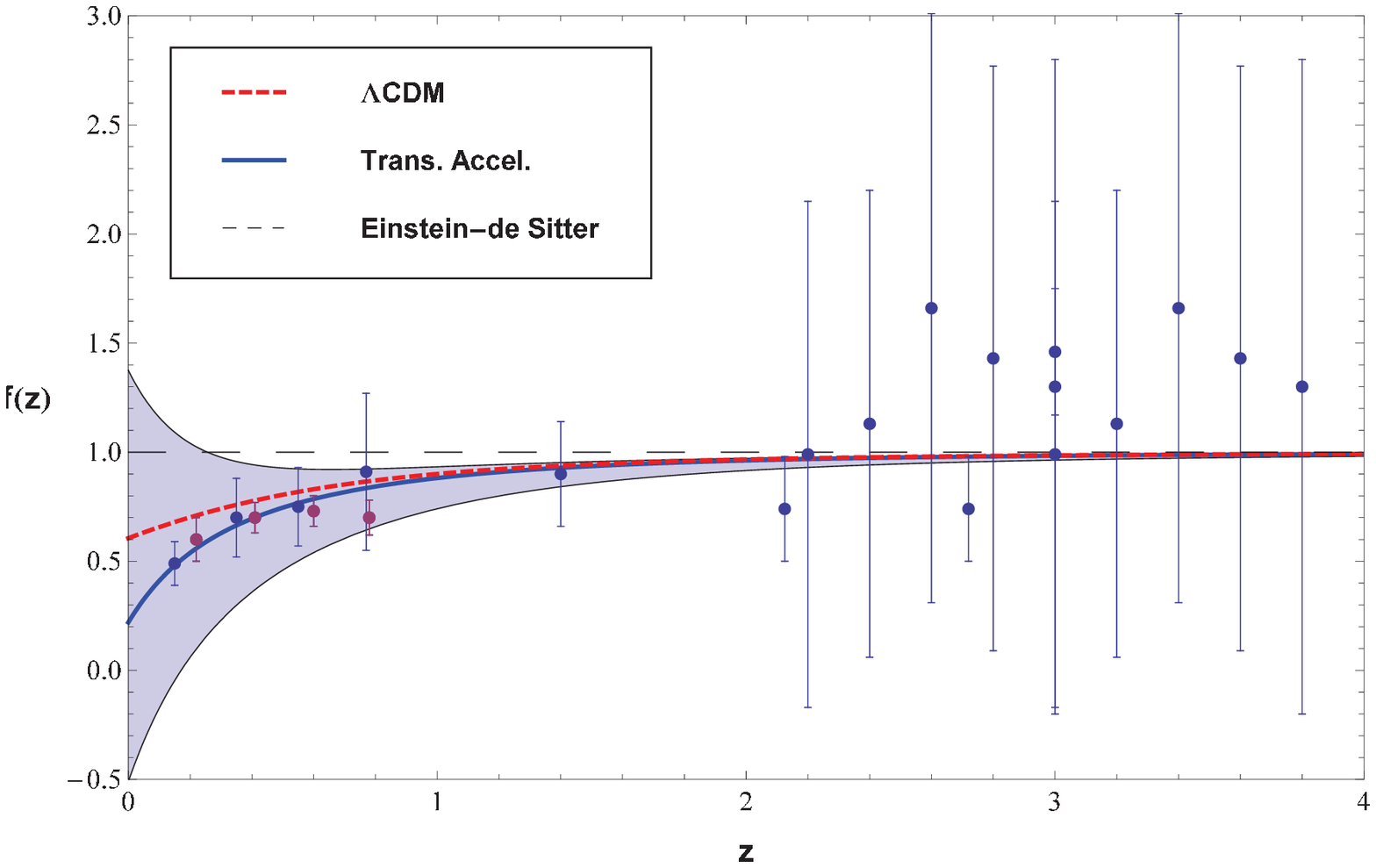,width=2.1in}}
\vspace*{8pt}
\caption{Dependence of the growth rate  $f(z)$ on the redshift $z$.\label{figgrowthf}}
\end{figure}

In a next step we consider the perturbation dynamics of this model.
Within a Newtonian approximation
it is convenient to introduce the growth  rate function
\begin{equation}
 f:= \frac{d \ln \delta_{m}}{d \ln a} \, , \label{gfunction}
\end{equation}
in terms of which the basic equation for $\delta_{m}$ takes the form
\begin{equation}
\frac{df}{d\ln a}\,  + \, f^{2} +  \left[a U(a) - 1\right]f = \frac{3}{2}\frac{G_{eff}}{G}\Omega_{m}
 \,,
\label{eqfGeff}
\end{equation}
where $G_{eff}$ is an effective gravitational constant which differs from $G$ due to the interaction terms.
Without interaction one has $a U(a) = 3/2$ as well as $G_{eff} = G$ and Eq.~(\ref{eqfGeff}) is equivalent
to Eq.~(\ref{deltaEdS}).
In Fig.~\ref{figgrowthf} the growth rate for the best-fit parameters of the present model is contrasted with the observations summarized in \cite{gong} as well as with those of \cite{blake}
and with the $\Lambda$CDM model.

\begin{figure}[ph]
\ \\
\ \\
\ \\
\centerline{\psfig{file=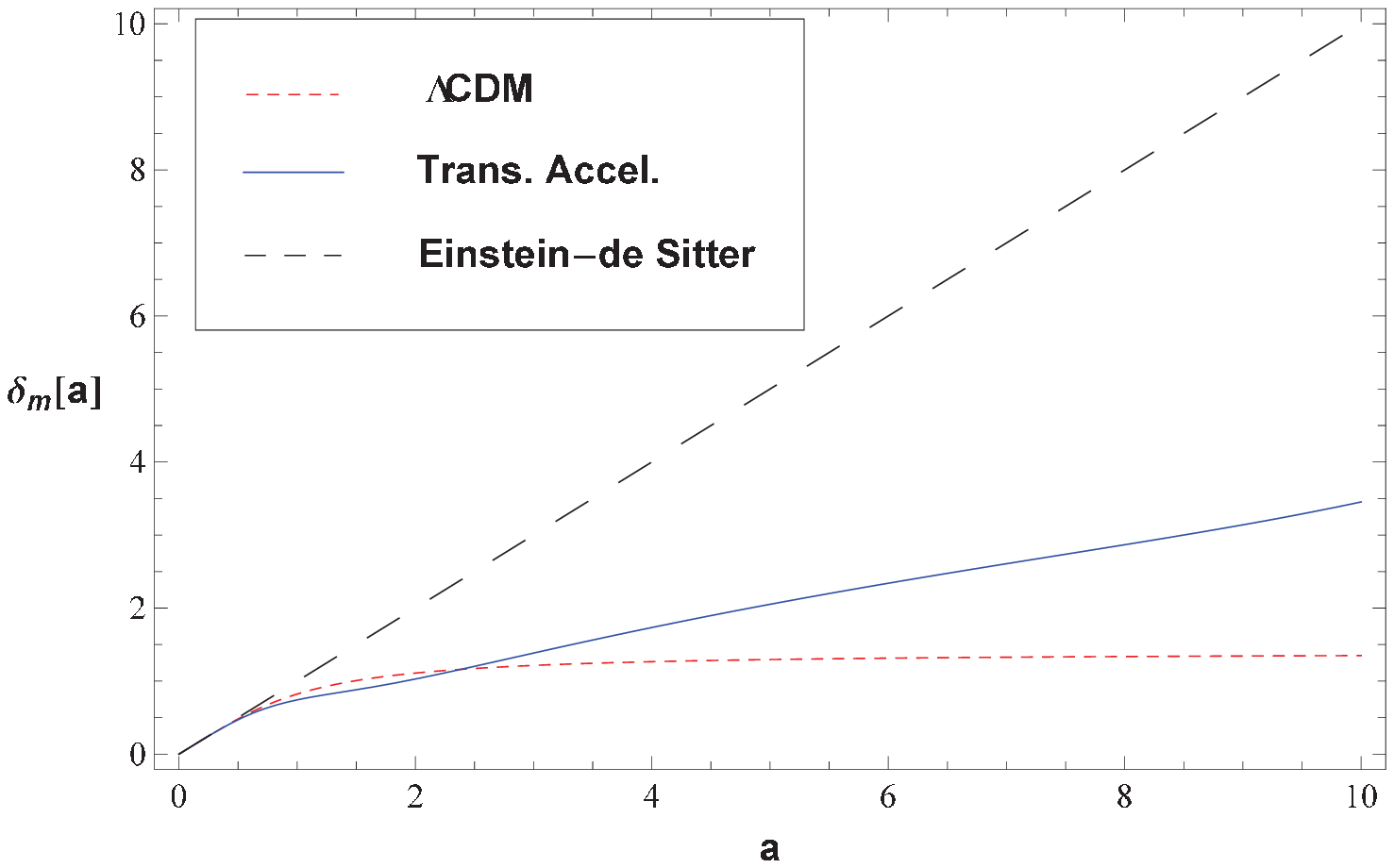,width=2.1in}}
\vspace*{8pt}
\caption{Fractional density perturbation as a function of the scale factor. Comparison between our best-fit model and the $\Lambda$CDM model. The straight line shows the corresponding increase for the Einstein-de Sitter universe.\label{figdelta1}}
\end{figure}

\noindent
Around the present epoch ($z\approx 0$) the deviation from the Einstein-de Sitter value is  larger  than that for the $\Lambda$CDM model,
corresponding to a slower growth of $\delta_{m}(a)$ for values of the order of $a\approx 1$.
This is seen in Fig.~\ref{figdelta1} as well,
which also shows predictions for a typical future behavior of $\delta_{m}(a)$. For $a>1$
the density contrast continues to grow in the transient acceleration scenario
while one has  $\delta_{m}(a) = $ const for the $\Lambda$CDM model.

In a relativistic perturbation theory the matter perturbation quantity has to be replaced by its
gauge-invariant counterpart according to (\ref{invc}). A general feature of dynamical DE models is the appearance of perturbations in the DE component itself, which are coupled to the perturbations of the matter density.
Although one expects that DE clumps less than DM, for the $\Lambda$CDM model the DE perturbations are
zero identically, it is not guaranteed from the outset that these perturbations can be neglected in a dynamical model \cite{Park-Hwang}.
To simplify the generally coupled system of equations for the density contrasts, we assume a proportionality between the gauge-invariantly defined fractional DE perturbations $\delta_{x}^{c} = \hat{\rho}_{x}^{c}/\rho_{x}$ and the matter perturbations $\delta_{m}^{c}$,
\begin{equation}\label{}
\delta_{x}^{c} = \epsilon \delta_{m}^{c}\,.
\end{equation}
The parameter $\epsilon$ quantifies the relative magnitude of the perturbations of the DE.
We then end up with a perturbation equation
\begin{equation}
\delta_{m}^{c\prime\prime} + F(a)\delta_{m}^{c\prime} + G(a) \delta_{m}^{c} = 0
\ ,
\label{deltaprprrel}
\end{equation}
where $F(a)$ and $G(a)$ are entirely determined by the analytically known background dynamics and
$G(a)$ depends on the scale of the perturbation \cite{cristofherpert}.
In the absence of interactions we recover the Einstein-de Sitter limit (\ref{deltaEdS})
of equation (\ref{deltaprprrel}).

Our aim is to calculate the matter power spectrum, defined in (\ref{Pk}).
To choose appropriate initial conditions, we use the circumstance that
at early times, i.e. for small scale factors $a \ll 1$, the equation (\ref{deltaprprrel}) has the asymptotic Einstein-de Sitter form (\ref{deltaEdS}),
which also coincides with the corresponding equation of the $\Lambda$CDM model at that period.
This allows us to relate our interacting model to the $\Lambda$CDM model at high redshift.
We shall benefit from the fact that the matter power spectrum for the
$\Lambda$CDM model is well fitted by the BBKS transfer
function \cite{bbks}.
Integrating the $\Lambda$CDM model back from today to a distant past, say $z = 10^{5}$, we
find the shape of the transfer function at that moment.
The obtained spectrum is then used as initial
condition for our model. This procedure was described in
more detail in references \cite{sola,sauloIC}.

\begin{figure}[ph]
\ \\
\ \\
\ \\
\ \\\
\centerline{\psfig{file=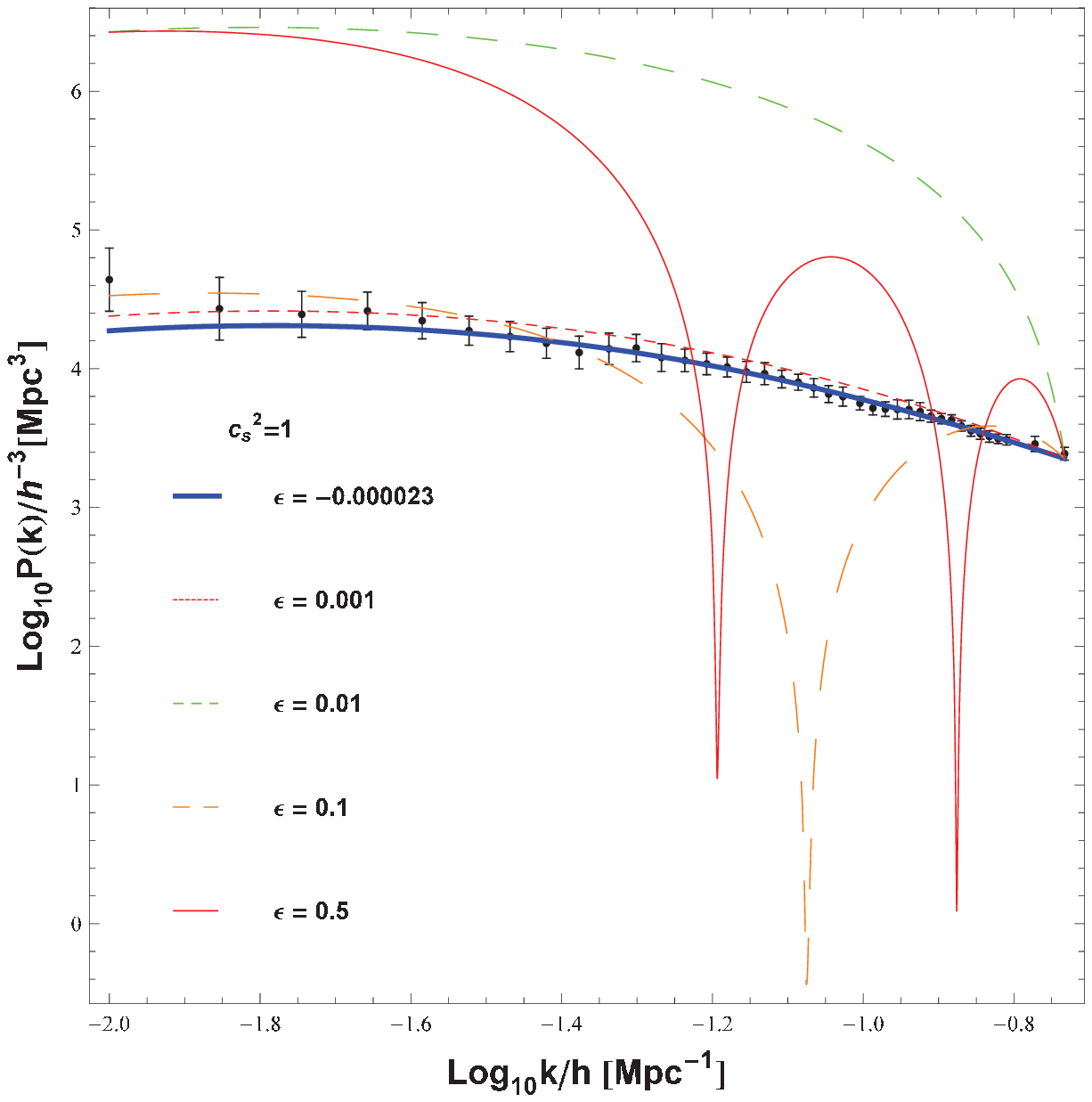,width=1.7in}}
\vspace*{8pt}
\caption{Matter power spectrum for $c_{s}^{2} = 1$ and different values of $\epsilon$. The thick solid (blue) curve ($\epsilon = -0.000023$) represents the best overall fit. On large scales, however, the curve with $\epsilon = 0.001$ shows the better performance. Larger values of $\epsilon$ result in (non-observed) oscillations. Data from 2dFGRS.\label{figespectro1+}}
\end{figure}

In Fig.~\ref{figespectro1+} we display the power spectrum, based on the 2dFGRS data \cite{cole}, for different values of $\epsilon$ for $c_{s}^{2} = 1$.
Here, $c_{s}^{2}$ is the square of the sound speed in the rest frame, defined by $\hat{p}^{c}_{x} = c_{s}^{2}\hat{\rho}_{x}^{c}$.
The scale dependence via the coefficient $G(a)$ in eq.~(\ref{deltaprprrel}) is sensitive to the product $\epsilon c_{s}^{2}$. The thick (blue) curve ($\epsilon = -0.000023$) represents the best overall fit. On large scales, however, the curve with $\epsilon = 0.001$ shows the better performance.
Obviously, only a very small factor $\epsilon$ is compatible with the data. Otherwise, there appear unobserved oscillations in the matter power spectrum which
are similar to those in (generalized) Chaplygin gases \cite{Sandvik,vdf}.
Although a constant $\epsilon$ corresponds to a very rough approximation, these results indicate that fluctuations of the DE component are small indeed on scales that are relevant for galaxy formation.

\begin{figure}[ph]
\ \\
\ \\
\ \\
\centerline{\psfig{file=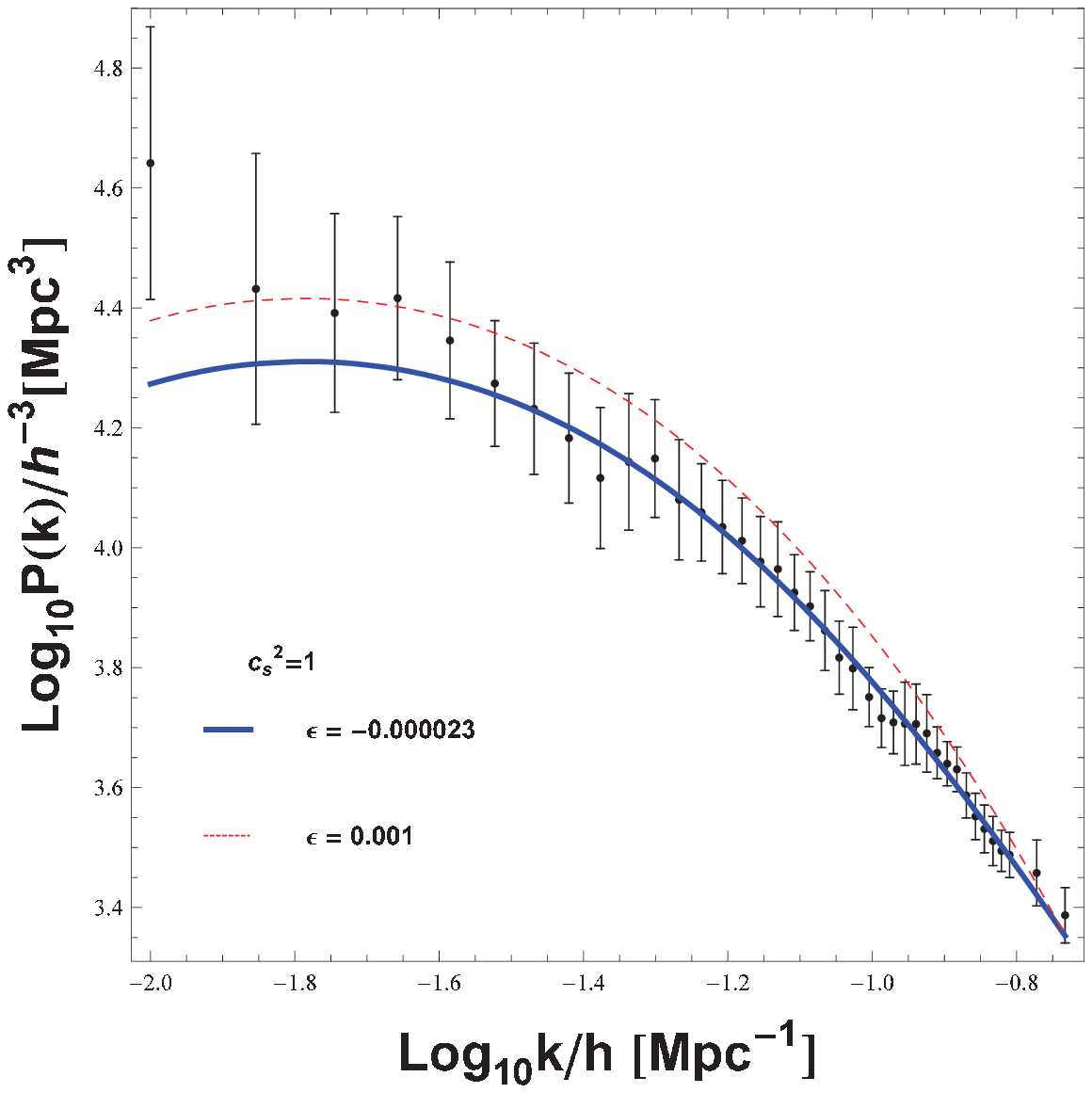,width=1.7in}}
\vspace*{8pt}
\caption{Matter power spectrum for $c_{s}^{2} = 1$.
While the total best-fit value for $c_{s}^{2} = 1$ is $\epsilon = - 0.000023$, it is obvious that on larger scales the dashed curve with $\epsilon = 0.001$ gives a better description. This corresponds to
the expectation that DE perturbations are more relevant on the largest scales.\label{Spectrum32}}
\end{figure}

As Fig.~\ref{Spectrum32} shows, even a very small value of $\epsilon$, although considerably larger than the best-fit value, influences the spectrum substantially on larger scales.
This indicates an increasing role of the DE perturbations with increasing scale. Consequently, for a more advanced analysis a scale-dependent $\epsilon$ should be used.

Finishing this subsection, we recall the basic general feature of this model. It is the double role of the interaction, which, on the one hand has to cancel a bare cosmological constant and, at the same time, it has to generate a current accelerated expansion. The coincidence problem reappears in a modified manner.
In the context of this model it would amount to the question, why the interaction strength is of the appropriate order to trigger an accelerated expansion just at the present epoch. 

\subsection{Decaying vacuum energy}
\begin{figure}[ph]
\ \\
\centerline{\psfig{file=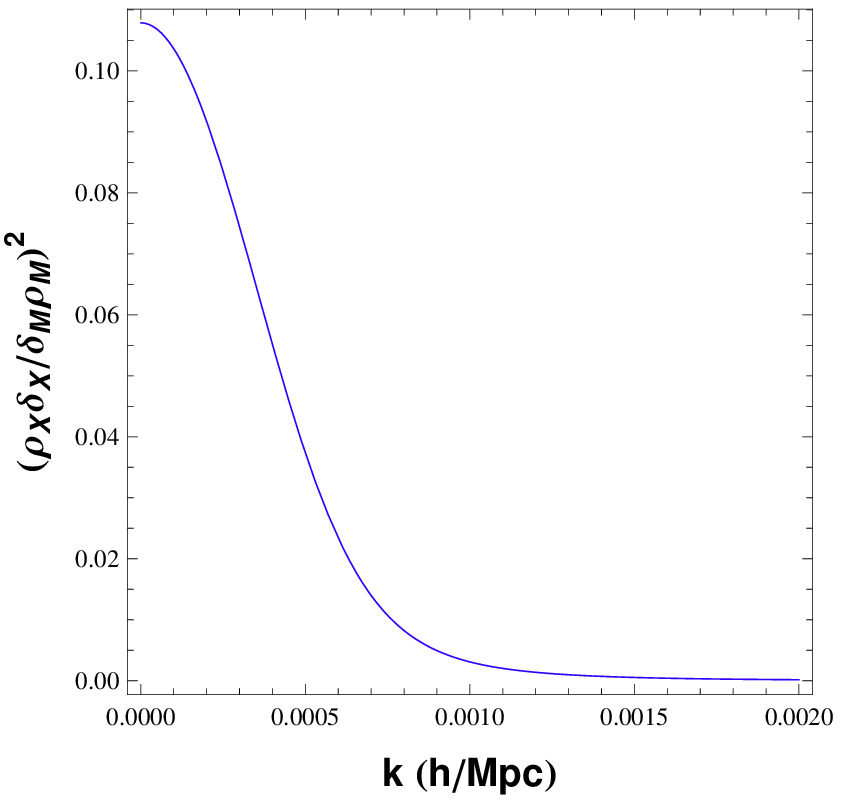,width=1.7in}}
\caption{Relative power spectrum as a function of $k$ on large scales for $\Omega_{m0} = 0.3$.\label{dXdM}}
\end{figure}

The dynamical DE scenario to be discussed in this subsection is based on a prescribed decay of the cosmological term, interpreted as vacuum energy.
The only preferred time scale in a homogeneous and isotropic universe is the Hubble time
$H^{-1}$. It is therefore tempting to associate a supposed vacuum decay with this scale.
The simplest case is a linear dependence $\rho_{X} \propto H$.
This dependence has some support from QCD \cite{QCD}, but it is treated here as a phenomenological approach \cite{humberto}.
Written in a covariant manner for later use in perturbation theory, the DE density of this model is characterized by
\begin{equation}\label{rhodecvac}
\rho_{x} = \frac{\rho_{0}}{3H_{0}}\left(1 - \Omega_{m0}\right)\Theta\ ,\qquad p_{x} = - \rho_{x}\,.
\end{equation}
In the background $\Theta = H/3$ is valid, where
the Hubble rate $H$ for this model, which has no $\Lambda$CDM limit, is
\begin{equation}\label{Hdec}
H = H_{0}\left[1 - \Omega_{m0} + \Omega_{m0}a^{-3/2}\right]\ .
\end{equation}
The coincidence problem is alleviated in so far as the energy-density ratio scales as
$a^{-3/2}$,
\begin{equation}\label{ratiodec}
\frac{\rho_{m}}{\rho_{x}} = \frac{\Omega_{m0}}{1 - \Omega_{m0}}a^{-3/2}\ ,
\end{equation}
compared with the $a^{-3}$ behavior of the $\Lambda$CDM model \cite{zimdahl}.
The perturbation dynamics is governed by an equation of the structure of (\ref{deltaprprrel}).
It is a particular feature of this model that it allows us to calculate explicitly the perturbations of the DE
component. The latter is determined by the combination
\begin{equation}
 \delta_{x}^{c}= - \frac{1}{3J} \left(a\delta^{c\prime}_{m} + B \delta_m^{c} \right)\,
\end{equation}
where
\begin{equation}
J = 1 + \frac{A}{3} \left[ 1 - \frac{B}{2} - \frac{A}{3}\,\frac{k^{2}}{a^{2}H^{2}}\right]
\
\label{K1}
\end{equation}
with the comoving wavenumber $k$ and
\begin{equation}
A = \frac{1 - \Omega_{m0}}{\ \Omega_{m0}a^{-3/2}}
\ ,\qquad
B = \frac{1 - \Omega_{m0}}{1 - \Omega_{m0} + \Omega_{m0}a^{-3/2}}
\ .
\label{s/H}
\end{equation}
Through the last term in (\ref{K1}) the factor $J$ and hence the DE perturbations are explicitly scale
dependent. For sub-horizon scales $\frac{k^{2}}{a^{2}H^{2}} \gg 1$ it follows that $|J| \gg 1$ and, consequently, $|\delta_{x}| \ll |\delta_{m}|$.
As shown in Fig.~\ref{dXdM}, only on the very largest scales DE perturbations may become noticeable.

This model of a decaying cosmological ``constant" was shown to be equivalent to a scenario in which
DM particles are created at a constant rate \cite{saulo12}. To be consistent with observations, a value
of $\Omega_{m0}$ of the order of $\Omega_{m0} \approx 0.45$ is required.

\subsection{Nonlinear interactions}
In this subsection we consider nonlinear interactions between DM and DE and
demonstrate, in the context of a dynamical system analysis, that such coupling may result in a future evolution of the Universe which differs from that of the standard $\Lambda$CDM model \cite{annap}.
Our starting point is again the system (\ref{balbg}). With $r = \rho_{m}/\rho_{x}$ and
$Q \equiv - 3H \Pi$, where $\Pi$ enters as an effective pressure,
the basic system for the two-component dynamics then is
\begin{equation}\label{}
\frac{d \rho}{d \ln a} = - \left(1 + \frac{w}{1+ r}\right)\rho\ , \qquad
\frac{d r}{d \ln a} = r \left[w - \frac{\left(1 + r\right)^{2}}{r \rho}\,\Pi\right]\,.
\end{equation}
There are stationary solutions (subscript st)
\begin{equation}\label{}
r_{st} = - 1 - w\ ,\qquad \rho_{st} = - \frac{w}{1+w}\Pi_{st}\,,
\end{equation}
where $\Pi_{st} = \Pi_{st}(\rho_{st}, r_{st})$ and $\Pi \neq \alpha\rho$.
Since $r_{st} > 0$, one has necessarily $w < -1$ and $\Pi_{st} < 0$, i.e., an EoS of the phantom type and $Q > 0$, i.e., an energy transfer from DE to DM. Since the individual energy densities $\rho_{m}$ and $\rho_{x}$ in terms
of $\rho$ and $r$ are
\begin{equation}\label{}
    \rho_{m} = \frac{r}{1+r}\rho\ \quad \mathrm{and }\quad \rho_{x} = \frac{1}{1+r}\rho\,,
\end{equation}
respectively, it seems convenient to rely on an ansatz
\begin{equation}\label{ans}
\Pi = - \gamma \rho^{m}r^{n}\left(1+r\right)^{s} = - \gamma\rho^{m+s}\rho_{m}^{n}\rho_{x}^{s-n}
\end{equation}
for the effective pressure term $\Pi$. Such a structure allows us to recover
the following known (linear) interaction models as special cases:
the choice $(m,n,s)=(1,0,-1)$ gives rise to an interaction $Q = 3\gamma H\rho_{x}$ and for
$(m,n,s)=(1,1,-1)$ one has $Q = 3\gamma H\rho_{m}$ (see, e.g., \cite{roy,abdalla}).
The structure (\ref{ans}) also contains the analytically solvable nonlinear model
$(m,n,s)=(1,1,-2)$ for which $Q = 3H\gamma\frac{\rho_{m}\rho_{x}}{\rho}$ with the
solution
\begin{equation}\label{}
r = r_{0} a^{3\left(w + \gamma\right)}\ , \qquad \rho = \rho_{0} a^{-3\left(1+w\right)}\left[\frac{1+r_{0} a^{3\left(w + \gamma\right)}}{1+r_{0}}\right]^{\frac{w}{w +\gamma}}\,.
\end{equation}
This solution coincides with the previously discussed scaling solution, based on the ansatz (\ref{rxi}) which becomes manifest if we identify $\gamma = - \left(w + \frac{\xi}{3}\right)$.

For an arbitrary combination of the parameters $m$, $n$ and $s$, analytical solutions of the nonlinear system are hardly available.
To get insight into the behavior of the system under more general conditions we shall resort here to a dynamical system analysis.
This analysis is based on the circumstance that, close to the critical points, the (generally unknown) solution of the nonlinear system behaves as the solution of the system, linearized around the critical points (Hartmann's theorem and, for purely imaginary eigenvalues, the Center Manifold Theorem (see, e.g., \cite{lynch} and \cite{boehmer}).
Using standard techniques, the characteristic equation for the critical points in our case is
\begin{equation}
\lambda_{\pm} = -\frac{1}{2}\left[2+ s + \left(1+n+s\right)w\right]
\left\{1
\mp\sqrt{1
+ \frac{4\left(m-1\right)\left(1+w\right)}{2+ s + \left(1+n+s\right)w}}\right\}
\ .
\label{lambdasol2}
\end{equation}
For $m\neq 1$ the general classification provides us with the following set of critical points:
\begin{itemize}
  \item Attractor
  for $m>1$ and $s < - (2+\left(1+n\right)w)/(1+w) - 2\sqrt{(1-m)/(1+w)}$\\
  \item Unstable for $m>1$ and $s > - (2+\left(1+n\right)w)/(1+w)  + 2\sqrt{(1-m)/(1+w)}$\\
  \item Saddle for $m< 1$, for all $n$ and $s$\\
  \item Center for $m>1$ and $2+ s + \left(1+n+s\right)w = 0$  for $n>1$\\
  \item Stable focus for $m>1$ and $2+ s + \left(1+n+s\right)w > 0$\\
  \item Unstable focus for $m>1$ and $2+ s + \left(1+n+s\right)w < 0$
\end{itemize}
\begin{table}[ht]
\tbl{Examples for an attractor as critical point.}
{\begin{tabular}{@{}ccccc@{}} \toprule
m\hphantom{00}&\hphantom{00}n & \hphantom{0}s  & Q & w \\
\colrule
$\frac{3}{2}$\hphantom{00} & \hphantom{0}$0$& \hphantom{0}$-1$ & $3H\gamma \sqrt{\rho}\rho_{x}$ & $-1.5 \leq  w < -1$\\
$\frac{3}{2}$\hphantom{00} & \hphantom{0}$\frac{1}{2}$& \hphantom{0}$-\frac{3}{2}$& $3H\gamma \sqrt{\rho_{m}}\rho_{x}$& $-1.125 \leq  w < -1$\\
$\frac{3}{2}$\hphantom{00} & \hphantom{0}$\frac{1}{2}$& \hphantom{0}$-1$ & $3H\gamma
\sqrt{\rho \rho_{x}\rho_{m}}$ & $-1.101 \leq  w < -1$\\
$2$\hphantom{00} & $\hphantom{0}\frac{1}{2}$& \hphantom{0}$-3$ & $3H\gamma \rho_{x}\sqrt{\rho\rho_{m}}$ & $-1.0625 \leq w<-1$\\ \botrule
\end{tabular}}
\end{table}
As an example we consider the attractor solution $(m,n,s)=(2,0,-2)$.
The critical points are
\begin{equation}\label{}
\rho_{st} = \frac{|w|\left(|w| - 1\right)}{\gamma} \,,\qquad r_{st} = |w|-1\,.
\end{equation}
The effective EoS parameter $w/(1+r)$ and the deceleration parameter $q$ approach their stationary values
through a power-law decay
\begin{equation}\label{}
\frac{w}{1+r} =  - 1 + \frac{g_{0}a^{3\lambda}}{|w|}\, \quad \mathrm{and} \quad q = - 1 + \frac{3}{2}\frac{g_{0}a^{3\lambda}}{|w|}\,,
\end{equation}
respectively, with the power
\begin{equation}\label{}
\lambda_{1,2} = - \frac{|w|}{2} \pm \sqrt{\frac{|w|^{2}}{4} - \left(|w| - 1\right)}\  < 0\,.
\end{equation}
Other examples for attractor solutions are listed in Table~3.

To summarize: the basic features of the dynamical system analysis for the interaction models
of this subsection are: i) necessarily a phantom-type EoS for the DE, ii) an energy transfer from
DE to DM and iii) the avoidance of a big-rip singularity due to the interaction.
For stable critical endpoints with finite values of the energy-density ratio $r$ the coincidence problem is obviously alleviated.

\section{Summary and Discussion}
Although the $\Lambda$CDM model \textit{grosso modo} is consistent with most observational data,
the study of alternative descriptions continues to be of interest.
Any competitive dynamical DE model has to make
predictions for the currently observed cosmic dynamics that are similar to those of the $\Lambda$CDM model.
We have reviewed here recent studies on interacting DE models.
Investigating models with interactions in the dark sector allows us to address the coincidence problem.
The problem of interacting models is to identify observational features which can unambiguously
be attributed to a certain coupling.
Interactions may provide corrections to uncoupled dark-sector models.
But there are also models for which the accelerated expansion is an interaction phenomenon.
Moreover, scenarios with nongravitational couplings in the dark sector may result in a
future evolution of the Universe which is different from that of the $\Lambda$CDM model.

\section*{Acknowledgments}
I thank the organizers of the 49th Winter School of Theoretical Physics, in particular
Zbigniew Haba, Andrzej Borowiec and Aneta Wojnar, for kind hospitality.
Special thanks to Fabiola Ar\'{e}valo, Humberto Borges, Saulo Carneiro, David R. Castro, J\'{u}lio Fabris, Wiliam Hip\'{o}lito-Ricaldi, Rodrigo vom Marttens, Diego Pav\'{o}n, Nelson Pinto-Neto, Anna Paula Ramos Bacalhau,  Hermano Velten and Cristofher Zu\~{n}iga Vargas, for enjoyable collaborations.
Support by CNPq and FAPES is gratefully acknowledged.

\end{document}